\documentstyle[aps,prl,twocolumn,psfig,floats]{revtex}

\begin{document}
\newcommand{\be}{\begin{eqnarray}}
\newcommand{\ee}{\end{eqnarray}}
\newcommand\del{\partial}

\wideabs{
\begin{flushright}
SUNY-NTG-02/30
\end{flushright}

\title  {Replica Limit of the Toda Lattice Equation}

\author {K. Splittorff and J.J.M. Verbaarschot}

 
\address{Department of Physics and Astronomy, SUNY, Stony Brook, New York 11794}

\maketitle
\begin  {abstract}
In a recent breakthrough Kanzieper showed that it is possible to obtain
exact nonperturbative Random Matrix results from the replica limit of
the corresponding Painlev\'e equation. In this article we analyze the
replica limit of the Toda lattice equation and obtain  exact
expressions for the two point function of the Gaussian Unitary Ensemble and the
resolvent of the  chiral Unitary Ensemble. In the latter case both
the fully quenched and the partially quenched limit are considered.
This derivation explains in a natural way the appearance of both compact
and noncompact integrals, the hallmark of the supersymmetric method, 
in the replica limit of the  expression for the resolvent. 
We also show that the supersymmetric
partition function and the partition function with fermionic replicas
are related through the Toda lattice equation.

\end {abstract}
}


\narrowtext

{\it Introduction.} 
The replica trick originally introduced in \cite{EA} has been
widely used in the theory of disordered systems
ranging from spin glasses \cite{EA,SK,Mezard} to QCD \cite{misha}. 
The quest for its rigorous understanding has been ongoing 
\cite{critique,KM,lerner,zirn,kanzieper,kim,DV1,ADDV,shinsuke,kanzieper02,Yan}
and will be further pursued in this letter.
The idea is to simplify the 
calculation of  the disorder average of the logarithm 
of the partition function starting from 
\be
\langle \log Z \rangle = \lim_{n \to 0} \left \langle \frac {Z^n -1}n
\right \rangle .
\ee
This method should work if we know the right hand side of this
equation as an analytical function of $n$. In practice, though, one
is usually only able to calculate $\langle Z^n\rangle$  for positive integer
values of $n$ which is not sufficient to reproduce the complete function.
For example, terms of the form $(\sin \pi n x)/n$ or singularities 
along the real $n$-axis could lead to a breakdown of the replica limit.
The first problems with the replica trick appeared in the theory 
of spin glasses \cite{SK} where the result for the entropy turned
out to be not positive definite. These problems were resolved by means
of an intricate scheme of replica symmetry breaking \cite{Mezard}.
Although the replica trick has been used widely in the theory of
disordered systems, it has been criticized as 
well \cite{critique,zirn}. In 
particular, it has been a long standing belief that the replica
trick can only be applied to asymptotic expansions which can be
resummed in some cases \cite{shinsuke}.
This belief was shattered in recent breakthrough
by Kanzieper \cite{kanzieper02}. In this letter 
we will obtain further
exact nonperturbative results from the replica limit of the Toda
lattice equation. Both this method and the method of \cite{kanzieper02}
rely on the relations between Random Matrix Theories and the theory
of exactly integrable systems. 
 
We will analyze the replica limit of correlation functions of
a Random Matrix partition function. In the case of the resolvent
we start from the identity
\be
G(x) = \frac 1N \left\langle {\rm Tr} \frac 1{x+D} \right\rangle=
\frac 1N \lim_{n\to 0}\partial_x \left\langle \frac {{\det}^n(D+x)}n
\right\rangle. 
\ee
Here, $D$ is an $N\times N$ random matrix and the average is over
the probability distribution of its matrix elements.
The replica  limit can be calculated in two different ways, by calculating
the partition function for fermionic replicas, i.e. for positive $n$, or
by using bosonic replicas, i.e. for negative $n$.

An exact method that avoids difficulties of 
the replica trick is the supersymmetric method \cite{Brezin,Efetov,VWZ}.
In that case the resolvent follows from the identity
\be
G(x) = \left .\frac 1N \partial_J \left 
\langle \frac {{\det}(D+x +J)}{{\det}(D+x )}
\right\rangle \right |_{J=0}.
\ee
This method is not applicable, though, to cases where the partition
function cannot be expressed as the average of a determinant,    
making it essential to understand the limitations of the replica limit.
In this letter we will not discuss other alternatives to the replica
method such as for example the dynamical method \cite{altland}.

Both the generating functions for the replica trick and
the supersymmetric method can be rewritten as a nonlinear 
$\sigma$-model.  Generally, it is believed 
\cite{EJ,VZR,critique,kim,ADDV} that  the replica
limit can only be applied to  
asymptotic expansions such
as the expansion in $1/x$ of the resolvent.
In \cite{critique} it was argued that the replica trick even fails to
reproduce the oscillatory 
terms in the asymptotic expansions  of  
correlation functions. Indeed, this is the case if one only takes
into account the leading saddle point manifold.  
The oscillatory terms in the asymptotic expansion can be obtained from
a subleading saddle point manifold \cite{KM}
(see also \cite{lerner,kanzieper}). However,  
exact results could only be derived when the asymptotic
expansion terminates \cite{zirn,DV1}. In \cite{zirn} this was related  
to the Duistermaat-Heckman localization property of the saddle point
integrals.

A real breakthrough was recently made by Kanzieper \cite{kanzieper02}. 
He succeeded to obtain exact nonperturbative results by means of
the replica method. His approach was based on the property that the
generating function for the resolvent or the two-point correlation function 
can be expressed in terms
of a solution of a Painlev\'e equation where the replica index
only occurs as a parameter in its coefficients. The 
replica limit of the resolvent is then obtained from solution of
this Painlev\'e equation for $n\to 0$. In particular, this allowed
him to reproduce exact nonperturbative results with an asymptotic
series that does not terminate. A first hint that it is possible to obtain
such exact results 
surfaced in \cite{DV1} where the leading
logarithmic singularity of the small $x$ expansion of the resolvent
was obtained from the replica trick.

In this letter we will analyze the replica trick for the two-point function
of the Gaussian Unitary Ensemble (GUE) and
for the microscopic limit of the resolvent
of the chiral Unitary Ensemble (chUE)
$-$ the random matrix ensemble relevant for QCD \cite{SV}.

The two-point function of the GUE can be expressed in terms of a
generating function for $n$ fermionic replicas 
\be
G(r) = -\lim_{n\to 0} \frac 1{n^2} \frac {\partial^2 Z_n(ir)}{\partial r^2}, 
\label{g(r)}
\ee
where the  eigenvalue representation of 
the nonlinear $\sigma$-model for the generating function is given by
 \cite{critique}
\be
 Z_n(ir) = n!\int_{-1}^{1}\prod_{k=1}^n d\lambda_k 
e^{-ir\lambda_k} \Delta^2(\{\lambda_k\}).
\label{zferm}
\ee  
Here, $\Delta(\{\lambda_k\})\equiv \prod_{k<l} (\lambda_k -\lambda_l)$ 
is the Vandermonde determinant.
For $n$ bosonic replicas, the two-point function is still given by
(\ref{g(r)}) but with a generating function $Z_{-n}(r)$, 
that now can be expressed as an integral over a noncompact manifold
\cite{critique},
\be
 Z_{-n}(ir) = \frac 1{n! [(n-1)!]^2}
\int_{1}^{\infty}\prod_{k=1}^n d\lambda_k 
e^{ir\lambda_k} \Delta^2(\{\lambda_k\}).
\label{zbos}
\ee  
The sign of the exponent is determined by the sign of the infinitesimal
increment of $ir$.  
The fermionic generating function satisfies 
the Toda lattice equation \cite{kanzieper}
\be
Z_n^2(ir)\partial_r^2 \log Z_n(ir) =-\frac {n^2}{(n+1)^2}
Z_{n+1}(ir) Z_{n-1}(ir).
\ee
If we interpret the generating function for negative $n$ as the 
the bosonic generating function (\ref{zbos}) this equation is valid
for all integer values of $n$. 
In the replica limit, the l.h.s. of this equation is $n^2$ times 
the connected two-point function. On the r.h.s. we  find from
(\ref{zferm}) and (\ref{zbos}) that
\be 
Z_1(ir) Z_{-1}(ir)= 2i \frac{\sin r}{ r} \frac {e^{ir}}r,
\ee
which is the analytical result for the two-point 
function \cite{critique}! In \cite{kanzieper} this result was
obtained from the replica limit of the Painlev\'e V equation. The present
method nicely explains the factorization of the two-point function into 
a compact and a non-compact contribution. In \cite{zirn} it was argued
that the replica limit reproduces the correct result in 
this case is because the asymptotic expansion in $1/r$ terminates. We
will now
analyze the microscopic limit of the resolvent $G(x)$ of the chUE 
for which the asymptotic expansion in $1/x$ does not terminate.
Nevertheless, we will obtain
the exact analytical result 
for $G(x)$ from the replica limit of
a Toda lattice equation.

For topological quantum number $\nu$  
the microscopic limit of the resolvent for the chUE is given by   
\be
G(x)=x(K_\nu(x)I_\nu(x)+K_{\nu-1}(x) I_{\nu+1}(x))
+\frac \nu x.
\label{valsup}
\ee
where $K_\nu(z)$ and $I_\nu(z)$ are modified Bessel functions. 
The most direct way to obtain this result is the supersymmetric method
\cite{OTV,DOTV,Yan} where the appearance of the compact and noncompact
integrals can be related to the symmetries of the underlying partition
function \cite{DV1}. However, it was first obtained in 
\cite{Vplb} by integrating the microscopic spectral density which can
be easily derived from the orthogonal polynomial method 
\cite{forrester,VZa,ADMN}.
In \cite{kanzieper02} this result was obtained by solving
the replica limit of the PIII Painlev\'e equation. In this approach,
the appearance of noncompact integrals in the replica limit is a
mystery. 
Here, we obtain the resolvent from the solution of the replica limit of 
the Toda lattice equation which relates the resolvent to both fermionic
and bosonic partition functions. The noncompact integrals then arise
from the bosonic partition function. This method is then generalized
to the partially quenched case, where the replica limit of a
generalized Toda lattice equation \cite{Kharchev} relates the
resolvent to a fermionic partition function and a partition function
with both fermions and bosons.
  
The partition function of the chUE
with $n$ fermionic replica flavors
with mass $m$ and $N_f$ additional fermionic 
flavors with masses $m_1, \cdots, m_{N_f}$ is defined by
\be
Z^{\rm chUE}_\nu = 
 \langle {\det}^n (D +m) \prod_{k=1}^{N_f}\det(D+m_k) \rangle .
\ee
The Dirac operator is given by
\be
D\equiv\left(\begin{array}{cc} 0 & W \\ W^\dagger & 0
\end{array}\right) ,
\ee
where $W$ is a rectangular $l\times(l+\nu)$ matrix so that $D$ has
exactly $\nu$ zero eigenvalues. The average is 
over the probability distribution of the matrix elements of $W$.
We will consider the microscopic limit
of this partition function where the thermodynamic limit is taken for 
fixed $x\equiv m\Sigma N$ and $x_k \equiv m_k \Sigma N$. Here,
$N=2l+\nu$ and $\Sigma=\lim_{m\to0,N\to\infty}\partial_m\log Z(m)/N$. In
this case the partition function
reduces to the unitary matrix integral \cite{SV}
\be
Z_{n,N_f}^\nu(x,\{x_i\}) = \frac 1{C_n}
\int_{U \in U(n+N_f) } \hspace{-12mm} dU {\det}^\nu U e^{\frac 12
{\rm  Tr}[M^\dagger U +M U^\dagger]},
\label{pq}
\ee
where $M={\rm
diag}(x,\cdots,x,x_1,\cdots,x_{N_f})$ is the mass matrix
and an additional normalization factor $C_n\! =\! \prod_{k=1}^n (k\!-1)!$
has been included for later
convenience. In the microscopic
limit, this 
partition function coincides with the partition function of QCD
\cite{LS,SV}. 
The replica limit of the resolvent is then given by 
\be
G(x,\{x_i\}) 
=\lim_{n \to 0}\frac 1n \partial_x \log Z_{n,N_f}^\nu(x,\{x_i\}). 
\ee
We first analyze the case of no additional massive flavors.

{\it The quenched resolvent.} In this case the partition function is given by
\be
Z_n^\nu(x) = \frac 1{C_n}
\int_{U \in U(n) } dU {\det}^\nu U e^{\frac x2{\rm  Tr}
[U +U^\dagger]}.
\ee
The unitary matrix integrals are well-known 
\cite{Brower,jsv,Tilo,baba}
\be
Z_n^\nu(x) = \frac 1{C_n}
\det [ I_{\nu+j-k}(x)]_{j=1,\cdots,n,\,\, k=1,\cdots, n}.
\label{fv}
\ee 
This expression is the $\tau$-function of an integrable hierarchy and 
satisfies the Toda lattice equation 
\cite{Okamoto,Kharchev,Witte,Poulpart}
\be
 \left [x\partial_x \right]^2\log Z_n^\nu(x) = nx^2
\frac{Z_{n+1}^\nu(x)Z_{n-1}^\nu(x)}{[Z_n^\nu(x)]^2}.
\label{toda}
\ee 
The choice of the normalization constants $C_n$
is consistent with the large $x$
asymptotic behavior of the l.h.s. of (\ref{toda})  which is given by $nx$.

In the replica limit, $n \to 0$, 
the partition function can be expanded as
\be
Z_n^\nu(x) =1 + n \sigma_1^\nu(x) + O(n^2).
\ee
In this limit, $Z_{-1}(x)$ appears 
in the r.h.s. of the
Toda equation, which can be interpreted as the partition function with
one bosonic replica. With normalization constant fixed by the large $x$
limit of (\ref{toda}) it is given by \cite{DV1}
\be
Z_{-1}^\nu(x) = 2K_\nu(x).
\ee
For the replica limit of the partition function we thus obtain the
differential equation
\be
n\left [x \partial x \right ]^2 \sigma_1^\nu(x)= 2n {x^2}
I_\nu(x) K_\nu(x).
\label{sig}
\ee 
Using the identity 
\be
\partial_x x^2 [ K_{\nu}(x) I_\nu(x)\! +\!
 K_{\nu-1}(x) I_{\nu+1}(x)]\! =\! 2 x K_\nu(x) I_\nu(x),
\label{id1}
\ee
 this differential equation can be integrated resulting in
\be
\sigma_1^\nu(x) = \int_0^{x} dt [t(K_\nu(t)I_{\nu}(t) +
K_{\nu-1}(t) I_{\nu+1}(t))
+\frac \nu t]
\ee
in agreement with the known result (\ref{valsup}) for the resolvent! 
The last term is a solution of the homogeneous differential equation
with integration constants 
fixed by the small $x$ expansion of the partition function.
 We notice that  
the discontinuity of the resolvent
arises through the bosonic integral $K_{\nu-1}(x)$ via the Toda
lattice equation. 
In this way, the analyticity in $x$ of 
the fermionic replica expression for the
resolvent \cite{zirn} is avoided.
In \cite{kanzieper02}, the expression for  
$ \sigma_1^\nu$,
was obtained from the solution of the replica limit of the PIII 
Painlev\'e equation. As a side remark, 
we notice that  $K_\nu(x)$ 
is a solution of the PIII equation for $n \to -1$ 
satisfying the boundary conditions for $n=-1$. 
The  PIII equation thus encodes the analyticity properties
of the bosonic integral $K_\nu(x)$.

{\it The partially quenched case.} The partition function (\ref{pq})
satisfies the recursion relation \cite{Kharchev}
\be
\label{genToda}
&& n x^2\frac{ Z^\nu_{n+1,N_f}(x,\{{x_k}\})
Z^\nu_{n-1,N_f}(x, \{x_k\})}
{[Z^\nu_{n,N_f}(x,\{x_k\})]^2} \\
 & = & 
\sum_{k=0}^{N_f} 
\left . x_0\partial_{x_0}x_k\partial_{x_k} 
\log Z^\nu_{n,N_f}(x_0,\{x_k\}) \right|_{x_0 =x} . \nonumber
\ee
This recursion relation can be derived from a special case of the Hirota
equation. For degenerate  masses $x_k$ the
sum over the derivatives is only over
the different masses. The normalization of the partition functions
is again fixed by the large $x$ behavior of the r.h.s. at fixed $x_k$ 
which is given by $nx$.

Let us consider the replica limit of (\ref{genToda}) for 
$N_f=1$. Then the numerator of the l.h.s. is the product  of
\be
Z^\nu_{1,1}(x, y) = \frac 2{y^2-x^2}\det \left |  \begin{array}{cc}
I_\nu(x) & I_\nu (y) \\
x I_{\nu+1}(x) &y I_{\nu+1} (y) 
\end{array} \right |,
\ee
 and the supersymmetric partition function \cite{DV1} 
\be
Z^\nu_{-1,1}(x,y) =
yK_\nu(x)I_{\nu+1}(y) + x
K_{\nu+1}(x) I_{\nu}(y)
\ee
with a bosonic mass
equal to $x$ and a fermionic mass equal to $y$. 
In the replica limit,  $Z^\nu_{n,1}(x,y)$ 
reduces to the partition function for one flavor and we have
\be
Z^\nu_{n,1}(x, y) = I_\nu(y)(1 + n \sigma_1^\nu(x,y) + O(n^2)).
\ee
Both sides of the recursion relation are of order $n$. The quantity 
$\sigma_1^\nu$ thus satisfies the linear differential equation
\be
(x\partial_x +y\partial y)x\partial_x \sigma_1^\nu(x,y) = x^2
\frac{Z^\nu_{-1,1}(x,y) Z^\nu_{1,1}(x,y)}{I_\nu(y)^2}.
\label{todanf}
\ee
Using (\ref{id1}) and the identities
\be
\label{ids}
&& (x\partial_x + y\partial_y)[ y I_\nu(x) I_{\nu+1} (y)
 - xI_\nu (y)  I_{\nu+1}(x)] \nonumber \\
&& =  (y^2-x^2)I_\nu(x) I_\nu(y),\nonumber \\
&&I_\nu^2(y)(x\partial_x + y\partial_y)\frac {K_\nu(x)}{I_\nu(y)} \nonumber \\
&& =  - [x K_{\nu+1}(x) I_\nu(y) + y K_\nu(x) I_{\nu+1}(y)] .
\ee
this equation
can be integrated with solution given by
\be
\sigma_1^\nu(x,y) & = & \int_0^x t
[I_\nu(t)K_\nu(t)+I_{\nu+1}(t)K_{\nu-1}(t) \nonumber \\
&& +\frac{\nu}{t^2}-\frac{K_\nu(t)}{I_\nu(y)} Z^\nu_{1,1}(t,y)] dt .
\ee
The resolvent given by $\partial_x \sigma_1^\nu(x,y)$ 
agrees with earlier results obtained by the 
supersymmetric method \cite{DOTV}.

The partially quenched resolvent  can be worked out analogously for more
flavors. For
example we have obtained the resolvent for
$N_f$ massless flavors and $\nu = 0$
from the replica limit of the Toda lattice equation. 

{\it General Solutions.} 
Away from the replica limit we can integrate the Toda lattice equation to
determine the partition functions with arbitrary number of fermions and
bosons but with some degenerate masses. Based on these results we conjecture
the general form of the partition function for arbitrary masses.
Up to a normalization constant we find for pure bosonic partition functions  
\be
\frac{\det[x^{j-1}_iK_{\nu+j-1}(x_i)]_{i,j=1,..,n}}
{\prod_{j>i}(x_j^2-x_i^2)}, 
\ee
which in the case of degenerate masses reduces to
$
Z_{-n,0}(x)=\det[K_{\nu+j-i}(x)].
$
This result agrees with the integral representation given in \cite{DV1,Yan}. 
The microscopic limit of the chUE partition function with $n$ bosons and
$m$ fermions satisfies the generalized Toda lattice equation
(\ref{genToda}) continued to negative $n$. Iterating this equation we
conjecture the following result for the partition function
\be
\label{Z-nm}
Z^\nu_{-n,m}(\{x_i\})\!=\!\frac{\det[x^{j-1}_i{\cal
J}_{\nu+j-1}(x_i)]_{i,j=1,..,n+m}}
{\prod_{j>i=1}^n(x_j^2-x_i^2)\prod_{ j>i=n+1}^{n+m}(x_j^2-x_i^2)}.  
\ee
The symbol ${\cal J}_\nu(x_i)\equiv(-1)^\nu K_\nu(x_i)$ for
$i=1,\ldots,n$ and ${\cal J}_\nu(x_i)\equiv I_\nu(x_i)$ for
$i=n+1,\ldots,n+m$. The structure of this expression is in agreement
with the general result obtained in \cite{FSgen}.
For $\nu=0$ the result for
$Z^\nu_{-n,m}$ was first derived by means of
the supersymmetric method \cite{FS}. 
The partition function (\ref{Z-nm}) is relevant for the calculation of
higher order spectral correlation functions \cite{DOTV,seif}.

{\it Conclusions.} 
The two-point function of the Gaussian Unitary Ensemble and the
quenched and partially quenched resolvent for the 
microscopic limit of the chiral Unitary Ensemble
have been obtained by solving
the replica limit of a (generalized) Toda lattice
equation. This method 
provides a direct analytical understanding of the 
appearance of compact and 
noncompact integrals in the expressions for the resolvent.
It also explains the appearance of  
a discontinuity in the fermionic replica limit of the resolvent.
The family of chUE partition functions with any number of fermions and
bosons and with arbitrary masses and topological charge also satisfies 
the generalized Toda lattice equation. On this basis we have conjectured a
form for this partition function that generalizes the result for zero
topological charge.

\noindent
{\bf Acknowledgements:} We wish to thank P.H. Damgaard and M.R. Zirnbauer
for useful discussion and suggestions.

\noindent
{\bf Note added:} The expression for $Z^\nu_{-n,m}(\{x_i\})$ in
eq. (\ref{Z-nm}) has now been verified \cite{FA} based on a
result in \cite{FSgen}.

\end{document}